\documentclass{iopconfser}

\usepackage[numbers,sort,compress]{natbib}

\usepackage{amsmath} % AMS Math Package
\usepackage{amssymb}	% Math symbols such as \mathbb
\usepackage{graphicx} % Allows for eps images
\usepackage{slashed}
\usepackage[colorlinks=true,urlcolor=blue]{hyperref}
\usepackage{xcolor}

\newcommand{\hbp}[1]{\hat{\bar{ #1 }}'}

\newcommand{\matTwo}[1]{\left(\begin{array}{cc} #1 \end{array}\right)}

\begin{document}

\title{Flipped $SU(5)$ GUT with conformal gravity from a single supermultiplet}

\author{David Chester$^{1}$, Alessio Marrani$^{2}$, Michael Rios$^{3}$ and Klee Irwin$^{1}$}

\affil{$^1$Quantum Gravity Research, Los Angeles, CA, USA}
\affil{$^2$Instituto de F{\'i}sica Teorica, Universidad de Murcia, Campus de Espinardo, Spain}
\affil{$^3$Dyonica ICMQG, Los Angeles, CA, USA}

\email{DavidC@QuantumGravityResearch.org (DC); jazzphyzz@gmail.com (AM); mfrios@gmail.com (MR); Klee@quantumgravityresearch.org (KI)}

\begin{abstract}
The Grassmann envelope is used to find the $\mathcal{N}=1$ `superquasiconformal' algebra in $D=10+1$. %superconformal algebra in $D=11+3$.
The adjoint representation of this algebra is found to contain $\mathfrak{su}_{2,2}\oplus \mathfrak{u}_{1}\oplus \mathfrak{su}_{5}$ as a submaximal subalgebra, giving a spectrum of conformal gravity with flipped $SU_{5}\times U_{1}$ GUT. Combining Yang-Mills theory with MacDowell-Mansouri gravity over the conformal group is found to recover the Einstein-Hilbert action with a cosmological constant. An action for the theory is presented, which contains the gauge bosons of gravity and GUT, three generations of fermions in an efficient manner, and a new Higgs sector. By using the superalgebra for the entire multiplet, the theory gauges a non-supersymmetric subalgebra without introducing superpartners.
\end{abstract}

% How do I cut to 8 pages?

% 1 page intro
% 1 page Grassmann envelope
% 1 page 3 gens from SU(3,2)
% 1 page conformal gravity
% 1 page fields
% 1 page action
% Half page conclusion

% No subsections
% Grab the necessary equations
% Fill in text
% Bolster with references at the end...

% Use AI to reformat the references...

%\tableofcontents

%\pagebreak

\section{Introduction}

While $SU_{5}$ grand unified theory (GUT) has been ruled out experimentally \cite{Georgi:1974sy,Takhistov:2016eqm,Bajc:2016qcc}, flipped $SU_{5}$ GUT is still experimentally viable \cite{BARR1982,ALLES1983,ANTONIADIS1987,Ellis:2018moe}. Supersymmetric (SUSY) flipped GUT has also been recovered from heterotic string theory, M-theory, and F-theory \cite{ANTONIADIS1988,Antoniadis:1989zy,2002Faraggi,Chung:2010bn,Chen2011}. On the other hand, gauge gravity theories have been studied extensively, also in presence of local SUSY \cite{PhysRevLett.38.739,Kaku:1977pa,Kaku:1977rk,Ferrara:1977ij}. This manuscript presents a theory of flipped $SU_{5}$ GUT with a quartic Higgs potential, no unobserved superpartners, and an unobserved vector Higgs field in addition to the ${\bf10}$ GUT Higgs scalar. This theory utilizes a Lie supergroup as the isometry (rather than isotropy) group of the matter sector, thus proposing a novel way to implement SUSY in gravity theories.

$\mathfrak{e}_{8}$ has been used in various approaches to GUT and string theory \cite{BarsGunaydinE8,Konshtein:1980km,OliveWest1983,Ong:1984ej,Buchmuller:1985rc,Barr:1987pu,Mahapatra:1988gc,Ellwanger:1990ss,Diaconescu:2000wy,Diaconescu:2003bm,Adler:2002yg,Adler:2004uj,Lisi:2007gv,Distler:2009jt,Pavsic:2008sz,Castro:2009zzd,Chen2011,Castro:2014wna,2014Douglas,Aranda:2020noz,Aranda:2020zms,Chester:2020nim,Manogue:2022yol,Green:1984sg,Dixon:1985jw,1987Campbell,Beasley:2008dc,Heckman:2009mn,Witten:1996mz,Horava:1996ma,Kutasov:1996fp,Martinec:1996wn}. Rather than gauging $\mathfrak{e}_{8}$, the theory presented herein gauges $\mathfrak{su}_{2,2}\oplus \mathfrak{u}_{1}\oplus \mathfrak{su}_{5}$, a submaximal subalgebra of $\mathfrak{e}_{8(-24)}$. By applying the Grassmann envelope (cf. e.g. \cite{Repovs-Zaicev} and Refs. therein) to $\mathfrak{e}_{8(-24)}$ \cite{Bars1979,Shestakov_1993,Truini:2020sjp,Truini:2020vlz}, a new Lie superalgebra is found and utilized in the matter sector of the theory. This Lie superalgebra, denoted as $\Gamma(\mathfrak{e}_{8(-24)})$, is the $\mathcal{N}=1$ $D=11+3$ superconformal algebra, or, equivalently, the $\mathcal{N}=1$ `superquasiconformal' Lie algebra in $D=10+1$, namely the $\mathcal{N}=1$ superization of the `quasiconformal' Lie algebra in $D=10+1$, in the sense introduced in \cite{GKN} (see also \cite{Gunaydin:2004md}). On the other hand, SUSY in $D=11+3$ has been previously introduced \cite{Bars:1997ug,Sezgin:1997gr,Rudychev:1997ui,Bars:1996ab,Bars:1997xb,Bars:2006dy,Bars:2010zw}, and it has been recently revisited and generalized by three of the authors \cite{Rios:2018lhc,Rios:2019rfc,Paduroiu:2021cvg}.

This manuscript is organized as follows. Section \ref{SUSY} provides the $\mathcal{N}=1$ $D=10+1$ superquasiconformal algebra and describes how this can be used to encode three generations of matter with a ${\bf128}$ spinor. Section \ref{theory} presents a complete action based on this superalgebra, which contains conformal gauge gravity with flipped $SU(5)$ GUT. Section \ref{conc} ends with our concluding remarks.

\section{The $\mathcal{N}=1$ superquasiconformal algebra in $D=10+1$ with three generations of fermions}
\label{SUSY}

The Grassmann envelope can be applied to arbitrary Lie algebras with a $\mathbb{Z}_2$ grading (see e.g. \cite{Repovs-Zaicev} and also \cite{Shestakov_1993}; for recent physical applications, see  \cite{Truini:2020sjp,Truini:2020vlz}). Remarkably, the exceptional Lie algebras are the only (simple) Lie algebras with spinor representations in their adjoint representation, when decomposed with respect to a suitable (maximal or submaximal) orthogonal subalgebra; to the best of our knowledge, the unique exception to this fact is provided by $\mathfrak{so}_{8}$, and it is a consequence of the $Spin_{8}$ triality. Quite recently, this inspired Truini to consider the Grassmann envelope of the compact real form $\mathfrak{e}_{8(-248)}$, denoted $\Gamma(\mathfrak{e}_{8(-248)})$. Here, we instead apply the Grassmann envelope to the minimally non-compact (quaternionic) real form $\mathfrak{e}_{8(-24)}$, containing (as a maximal and symmetric subalgebra) $\mathfrak{so}_{12,4}$, which can be regarded as the $D=11+3$ conformal algebra, but also as the `quasiconformal' algebra in $D=10+1$, in the sense introduced some time ago in \cite{GKN}. Under the reduction to its maximal subalgebra $\mathfrak{so}_{12,4}$, the Lie algebra $\mathfrak{e}_{8(-24)}$ gets $\mathbb{Z}_2$- graded, and decomposes as follows :
\begin{equation}
\mathfrak{e}_{8(-24)} \rightarrow \mathfrak{so}_{12,4}: \,\, {\bf248} = {\bf120} \oplus {\bf128},
\end{equation}
where $\mathbf{120}\equiv \mathfrak{g}_{0}$ is the adjoint representation of $\mathfrak{so}_{12,4}$, and $\mathbf{128}\equiv \mathfrak{g}_{1}$ is a (Majorana-Weyl) semispinor representation of $\mathfrak{so}_{12,4}$ itself.

For any $\mathbb{Z}_{2}$-graded Lie algebra $\mathfrak{g}=\mathfrak{g}_{0}\oplus \mathfrak{g}_{1}$, the Grassmann envelope (denoted as $\Gamma(\mathfrak{g})$) is defined as
\begin{equation}
\Gamma(\mathfrak{g}) := \mathfrak{g}_0 \otimes G_0 \oplus \mathfrak{g}_1 \otimes G_1 =:\mathfrak{g}_{\overline{0}} \oplus \mathfrak{g}_{\overline{1}},
\end{equation}
where $G_0$ and $G_1$ are the even and odd parts of a Grassmann algebra, with each odd generator being tensored with a unique Grassmann number. Thus, two generators $T_a$ and $T_b$ of $\mathfrak{e}_{8(-24)}$ get mapped to $T_a \otimes e_a$ and $T_b \otimes e_b$, respectively. This allows for the Lie brackets in $\mathfrak{e}_{8(-24)}$ to give rise to super-Lie brackets in $\Gamma(\mathfrak{e}_{8(-24)})$,
\begin{equation}
T_a\otimes e_a \circ T_b\otimes e_b = [T_a,T_b]\otimes e_a e_b = -(-1)^{ij} [T_b, T_a] \otimes e_b e_a = -(-1)^{ij} T_b\otimes e_b \circ T_a\otimes e_a, \label{supercomm}
\end{equation}
where the indices $i,j=0,1$ denote the degree of the generators within the aforementioned $\mathbb{Z}_{2}$ grading. Recently, the super-Jacobi identity for the super-Lie brackets defined above has been proven to hold by Truini \cite{Truini:2020sjp,Truini:2020vlz}.

In order to explicitly present supercommutation relations, we split the generators of $\mathfrak{e}_{8(-24)}$ into the generators $J_{\hbp{a},\hbp{b}}$ of $\mathfrak{so}_{12,4}$ and into the generators $T^{\hbp{\alpha}}$, transforming like a semispinor under $\mathfrak{so}_{12,4}$ itself (the range of the indices are $\hbp{a} = 1,2,\dots 16$ and $\hbp{\alpha} = 1,2,\dots,128$). Such generators get respectively mapped to the generators $M_{\hbp{a}\hbp{b}}$ and $Q^{\hbp{\alpha}}$ of $\Gamma(\mathfrak{e}_{8(-24)})$, satisfying the following supercommutation relations:
\begin{eqnarray}
\Gamma(\mathfrak{e}_{8(-24)}): \qquad \qquad [M_{\hbp{a}\hbp{b}}, M_{\hbp{c}\hbp{d}}] &=& \eta_{\hbp{b}\hbp{c}}J_{\hbp{a}\hbp{d}} - \eta_{\hbp{b}\hbp{d}} J_{\hbp{a}\hbp{c}} - \eta_{\hbp{a}\hbp{c}} M_{\hbp{b}\hbp{d}} + \eta_{\hbp{a}\hbp{d}}M_{\hbp{b}\hbp{c}}, \\
{[}M_{\hbp{a}\hbp{b}}, Q^{\hbp{\alpha}}] &=& \frac{1}{4} \left( \gamma_{\hbp{a}}\gamma_{\hbp{b}} - \gamma_{\hbp{b}}\gamma_{\hbp{a}}\right)^{\hbp{\alpha}}_{\,\,\, \hbp{\beta}} Q^{\hbp{\beta}}, \\
\{ Q^{\hbp{\alpha}}, Q_{\hbp{\beta}}\} &=& \left( \gamma^{\hbp{a}}\gamma^{\hbp{b}} - \gamma^{\hbp{b}}\gamma^{\hbp{a}}\right)^{\hbp{\alpha}}_{\,\,\, \hbp{\beta}} M_{\hbp{a}\hbp{b}}. \label{se8}
\end{eqnarray}
Note that the generators $Q_{\hbp{\alpha}}$ factorize out a Grassmann parameter via $T_{\hbp{\alpha}} = \theta_{\hbp{\alpha}} Q_{\hbp{\alpha}}$ (indices not summed over); this is actually the opposite of the procedure discussed by Bars and G{\"u}naydin in \cite{Bars1979}.

%In the langauge from G{\"u}naydin, $\mathfrak{so}_{12,4}$ is the quasiconformal algebra $\mathfrak{qconf}(\mathbb{R}\oplus J_2(\mathbb{O}))$ and $\mathfrak{e}_{8(-24)} = \mathfrak{qconf}(J_3(\mathbb{O}))$. This implies that $\Gamma(\mathfrak{e}_{8(-24)}$ is the minimal chiral superextension of $\mathfrak{qconf}(\mathbb{R}\oplus J_2(\mathbb{O}))$, making it a superquasiconformal algebra $\mathfrak{sqconf}(\mathbb{R}\oplus J_2(\mathbb{O}))$. Recall that $\mathbb{R}\oplus J_2(\mathbb{O})$ admits vector representations of $D=10+1$, the signature of M-theory. F-theory is known to stem from 12-dimensions, while Bars and Sezgin discussed $D=11+3$ superalgebras, followed by Bars' notion of S-theory in $D=11+2$. The Grassmann envelope $\Gamma(\mathfrak{e}_{8(-24)}$ is the $\mathcal{N}=1$ superconformal algebra in $D=11+3$, which is a non-classical Lie superalgebra not previously found in the literature \cite{DAuria:2000byu}. The minimal classical Lie superalgebras containing $\Gamma(\mathfrak{e}_{8(-24)})$ are
%\begin{equation}
%\Gamma \left( \mathfrak{e}_{8(-24)}\right) \subset \left\{
%\begin{array}{l}
%\underset{\text{dim}=\left( 7,114|256\right) }{\mathfrak{osp}\left(
%64,64|2\right) }, \\
%\underset{\text{dim}=\left( 16,384|256\right) }{\mathfrak{su}\left(
%64,64|1\right) \equiv \mathfrak{su}\left( 1|64,64\right) }, \\
%\underset{\text{dim}=\left( 16,384|256\right) }{\mathfrak{sl}_{\mathbb{R}%
%}\left( 128|1\right) \equiv \mathfrak{sl}_{\mathbb{R}}\left( 1|128\right) },%
%\end{array}%
%\right.
%\end{equation}%

In light of the (non-linear) realization of simple Lie algebras as
symmetries of a generalized light cone (defined by a quartic irreducible
norm), introduced in \cite{GKN} and named `quasiconformal' realization, $%
\mathfrak{so}_{12,4}$ and $\mathfrak{e}_{8(-24)}$ can be regarded as the
quasiconformal Lie algebras associated to the cubic Jordan algebras $\mathbb{%
R}\oplus J_{2}\left( \mathbb{O}\right) $ and $J_{3}(\mathbb{O})$,
respectively : $\mathfrak{so}_{12,4}=\mathfrak{qconf}\left( \mathbb{R}\oplus
J_{2}\left( \mathbb{O}\right) \right) $ and $\mathfrak{e}_{8(-24)}=\mathfrak{%
qconf}\left( J_{3}\left( \mathbb{O}\right) \right) $. This implies that the
Lie superalgebra defined by\footnote{$\Gamma \left( \mathfrak{e}%
_{8(-24)}\right) $ is a non-classical Lie superalgebra which, to the best of
our current knowledge, has never been considered in literature (see e.g.
\cite{DAuria:2000byu}).} $\Gamma \left( \mathfrak{e}_{8(-24)}\right) $,
which is the $\mathcal{N}=1$ chiral superization of $\mathfrak{so}_{12,4}$,
can be regarded as the $\mathcal{N}=1$ supersymmetric extension of $%
\mathfrak{qconf}\left( \mathbb{R}\oplus J_{2}\left( \mathbb{O}\right)
\right) $, namely as the $\mathcal{N}=1$ `superquasiconformal' Lie
algebra in $10+1$ space-time dimensions : $\Gamma \left( \mathfrak{e}_{8(-24)}\right) \simeq \mathfrak{sqconf}\left(
\mathbb{R}\oplus J_{2}(\mathbb{O})\right) $.

Since the (maximal) subalgebras of $\mathfrak{e}_{8(-24)}$ are classified, one can consider the subalgebras of $\Gamma(\mathfrak{e}_{8(-24)})$. In this work, we will focus on $\mathfrak{su}_{3,2}\oplus \mathfrak{su}_5$ which is a maximal (non-symmetric) subalgebra of $\mathfrak{e}_{8(-24)}$, as three of the present authors (DC, AM and MR) previously showed that this provides phenomenologically relevant representations. Under $\mathfrak{e}_{8(-24)}\rightarrow \mathfrak{su}_{3,2}\oplus \mathfrak{su}_{5}$, the adjoint of $\mathfrak{e}_{8(-24)}$ branches as follows :
\begin{equation}
{\bf248} = ({\bf24},{\bf1}) \oplus ({\bf1},{\bf24}) \oplus ({\bf5},{\bf10}) \oplus (\overline{{\bf5}},\overline{{\bf10}}) \oplus ({\bf10},\overline{{\bf 5}}) \oplus (\overline{{\bf 10}},{\bf 5}).
\end{equation}
%By taking $\mathfrak{e}_{8(-24)} \rightarrow \mathfrak{su}_{3,2} \oplus \mathfrak{su}_5 \rightarrow \mathfrak{su}_{2,2}\oplus \mathfrak{u}_1 \oplus \mathfrak{su}_5$, the following representations are found,
By continuing the branching and considering the maximal (symmetric)
subalgebra $\mathfrak{su}_{2,2}\oplus \mathfrak{su}_{5}\oplus \mathfrak{u}_{1}$ of $\mathfrak{su}_{3,2}\oplus \mathfrak{su}_{5}$, one obtains

\begin{eqnarray}
({\bf24},{\bf1}) &=& ({\bf15},{\bf1})_0 \oplus ({\bf1},{\bf1})_0 \oplus ({\bf4},{\bf1})_5 \oplus (\overline{{\bf4}}, {\bf 1})_{-5}, \nonumber \\
({\bf5},{\bf10}) &=& ({\bf4},{\bf10})_1 \oplus ({\bf1},{\bf10})_{-4}, \\
({\bf10},\overline{{\bf 5}}) &=& ({\bf6},\overline{{\bf5}})_2 \oplus ({\bf4},\overline{{\bf5}})_{-3}, \nonumber
\end{eqnarray}
where the subscripts here denote the charges under $\mathfrak{u}_{1}$.

Thus, one can consider the Grassmann envelope $\Gamma \left(
\mathfrak{su}_{3,2}\right) $, where the relevant $\mathbb{Z}_{2}$-grading is
the one determined by tha maximal (symmetric) Lie subalgebra $\mathfrak{su}%
_{2,2}\oplus \mathfrak{u}_{1}\subset \mathfrak{su}_{3,2}$.
Since $\mathfrak{su}_{2,2}\simeq \mathfrak{so}_{4,2}$ is the conformal
algebra in $3+1$ space-time dimensions, and
%The Grassmann envelope of $\mathfrak{su}_{3,2} \subset \mathfrak{e}_{8(-24)}$ must also be a superalgebra, as it contains eight spinor degrees of freedom when breaking to $\mathfrak{su}_{2,2}\oplus \mathfrak{u}_1$. Remarkably, $\Gamma(\mathfrak{e}_{8(-24)})$ contains $\Gamma(\mathfrak{su}_{3,2})$ as the $\mathcal{N}=1$ superconformal algebra in $D=3+1$, since
\begin{eqnarray}
\mathfrak{su}_{3,2} &\rightarrow& \mathfrak{su}_{2,2}\oplus \mathfrak{u}_1, \\
{\bf 24} &=& {\bf4}_{5} \oplus \left( {\bf 15}_0 \oplus {\bf1}_0 \right) \oplus \overline{{\bf4}}_{-5}, \nonumber
\end{eqnarray}
$\Gamma \left( \mathfrak{su}_{3,2}\right) $ can be regarded as the $\mathcal{N}=1$ superconformal Lie superalgebra in $3+1$ space-time dimensions, with even and odd components given by the following $\mathfrak{u}_{2,2} $-modules
%The superalgebra $\Gamma(\mathfrak{su}_{3,2})$ contains the following even and odd parts
\begin{equation}
\mathfrak{g}_{\overline{0}} = {\bf15}_0 \oplus {\bf1}_0, \qquad \mathfrak{g}_{\overline{1}} = {\bf4}_5 \oplus \overline{{\bf 4}}_{-5},
\end{equation}
%This motivates a natural way to recover the $\mathcal{N}=1$ $D=3+1$ superconformal group from the $D=12+4$ $\mathcal{N}=1$ superconformal algebra.
and
the above reasoning provides a quite natural realization of the (non-maximal) embedding of the
$\mathcal{N}=1$ superconformal algebra in $3+1$ into the $\mathcal{N}=1$
superquasiconformal Lie algebra in $10+1$ dimensions :  $\Gamma \left(
\mathfrak{su}_{3,2}\right) \subset \Gamma \left( \mathfrak{e}%
_{8(-24)}\right) $, or, equivalently, $\mathfrak{sconf}\left( J_{2}(\mathbb{C})\right) \subset \mathfrak{sqconf}%
\left( \mathbb{R}\oplus J_{2}(\mathbb{O})\right) $.

Furthermore, Kostant demonstrated that $Spin_{4,4}$ intersects with $Sp_{8}(\mathbb{R})$ to give three charts of $U_{2,2}=SU_{2,2}\times U_{1}$, which was motivated to be relevant for three generations of matter \cite{KostantSO44a}. While $\mathfrak{e}_{8(-24)}$ does contain $\mathfrak{sp}_{8,\mathbb{R}}$, $Spin_{4,4} \cap SU_{3,2}$ also contains three charts of $SU_{2,2}\times U_{1}$.

Since $\mathfrak{so}_{4,4} \subset \mathfrak{so}_{12,4} \subset \Gamma(\mathfrak{e}_{8(-24)})$, it is possible to recover three generations of the standard model fermions from the ${\bf128}$ spinor, even though the standard model typically requires 192 off-shell degrees of freedom with $D=3+1$ spinors (when including right-handed neutrinos). This is because the extra time dimensions provide for an efficient way to account for mass/flavor oscillations by rotating between the different conformal charts. Three sets of projection operators acting on the ${\bf128}$ spinor are found to obtain three sets of spinors in $D=3+1$ by using the Clifford algebra $Cl(12,4)$, which also contains $E_{8(-24)}$ since the bivectors give $Spin_{12,4}$ and a minimal irreducible representation of $Cl(12,4)$ can identify a ${\bf 128}$ spinor. For simplicity, we demonstrate how three separate sets of ${\bf 16}$ spinors from $Spin_{10}$ GUT could be found, which is known to break to $SU_{5}\times U_{1}$.

Following Ref.~\cite{Floerchinger:2019oeo}, a basis of $Cl(12,4)$ is found from recursive tensor products of $Cl(0,2)$, $Cl(1,1)$ and $Cl(2,0)$.
\begin{eqnarray}
Cl(0,2): &&\quad  e^1 \equiv \gamma_{(0,2)}^1 = \matTwo{0&1\\1&0}, \quad e^2 \equiv \gamma_{(0,2)}^2 = \matTwo{1&0\\0&-1}, \nonumber \\
Cl(1,1): &&\quad  E^0 \equiv \gamma_{(1,1)}^0 = \matTwo{0&-1\\1&0}, \quad E^1 \equiv \gamma_{(1,1)}^1 = \matTwo{0&1\\1&0}, \\
Cl(2,0): &&\quad e^{-1} \equiv \gamma_{(2,0)}^{-1} = \matTwo{0&-i\\-i&0}, \quad e^{0} \equiv \gamma_{(2,0)}^{-2} = \matTwo{0&-1\\1&0} .\nonumber
\end{eqnarray}
Refs.~\cite{Floerchinger:2019oeo} and \cite{DAuria:2000byu} differ in conventions. A basis for $Cl(4,12)$ is identified as
\begin{eqnarray}
Cl(4,12) &=& (Cl(2,0) \otimes Cl(0,2))^{\otimes 2} \otimes (Cl(1,1))^{\otimes 4}, \nonumber \\
&&\begin{array}{rll}
\Gamma^{-3} =& I_{(0,8)}\otimes E^0\otimes(E^{01})^{\otimes 3} , & \quad \Gamma^{5} = I_{(0,4)}\otimes e^{-1}\otimes e^{12}\otimes (E^{01})^{\otimes 4}, \\
\Gamma^{-2} =& I_{(1,9)}\otimes E^0\otimes(E^{01})^{\otimes 2}, & \quad \Gamma^{6} = I_{(0,4)}\otimes e^{0}\otimes e^{12}\otimes (E^{01})^{\otimes 4}, \\
\Gamma^{-1} =& I_{(2,10)}\otimes E^0\otimes E^{01}, & \quad \Gamma^{7} = I_{(6,0)}\otimes e^{1}\otimes (E^{01})^{\otimes 4}, \\
\Gamma^{0} =& I_{(3,11)}\otimes E^0, & \quad \Gamma^{8} = I_{(6,0)}\otimes e^{2}\otimes (E^{01})^{\otimes 4}, \\
\Gamma^{1} =& \gamma^{-1}_{(2,0)}\otimes e^{12}\otimes e^{-10}\otimes e^{12}\otimes (E^{01})^{\otimes 4} , & \quad \Gamma^{9} = I_{(0,8)}\otimes E^{1}\otimes (E^{01})^{\otimes 3}, \\
\Gamma^{2} =& \gamma^{0}_{(2,0)}\otimes e^{12}\otimes e^{-10}\otimes e^{12}\otimes (E^{01})^{\otimes 4}, & \quad \Gamma^{10} = I_{(1,9)}\otimes E^{1}\otimes (E^{01})^{\otimes 2}, \\
\Gamma^{3} =& I_{(2,0)}\otimes e^{1}\otimes e^{-10}\otimes e^{12}\otimes (E^{01})^{\otimes 4}, & \quad \Gamma^{11} = I_{(2,10)}\otimes E^{1}\otimes E^{01}, \\
\Gamma^{4} =& I_{(2,0)}\otimes e^{2}\otimes e^{-10}\otimes e^{12}\otimes (E^{01})^{\otimes 4}, & \quad \Gamma^{12} = I_{(3,11)}\otimes E^1,
\end{array}
\end{eqnarray}
where $e^{12} = e^1 e^2$, $E^{01} = E^{0} E^1$, and $e^{-10} = e^{-1} e^0$ and $I_{p,q}$ is the $2^{(p+q)/2} \times 2^{(p+q)/2}$ identity matrix.

Generalizing from Ref.~\cite{Chester:2020nim}, each of the three conformal charts have a unique projection operator $P_i^N$ for $i=1,2,3$ to obtain normal matter and remove mirror fermions,
\begin{eqnarray}
P_1^N &=& \frac{1}{2} \left( I_{(12,4)} + \Gamma^{-14}\right), \nonumber \\
P_2^N &=& \frac{1}{2} \left( I_{(12,4)} + \Gamma^{-24}\right), \\
P_3^N &=& \frac{1}{2} \left( I_{(12,4)} + \Gamma^{-34}\right). \nonumber
\end{eqnarray}
Since the gamma matrices for $Cl(12,4)$ are $256\times 256$, the ${\bf 128}$ Majorana-Weyl spinor are embedded in a 256-dimensional bispinor $\Psi$ with 128 components set to zero. The projection operators identify three left-chiral spinors $\psi_i \equiv P_{i}^N \Psi$. Each $\psi_i$ contains 64 off-shell degrees of freedom with 32 on-shell degrees of freedom. While the 96 on-shell degrees of freedom are independent, the 192 off-shell degrees of freedom overlap and contain 128 independent degrees of freedom.

$D=3+1$ Weyl spinors require complex structure with an imaginary unit and a complex conjugation operator. Three sets of complex units are found as rotations in the extra time dimensions of the Clifford algebra, giving the quaternionic units $i=\Gamma^{-2-3}$, $j=\Gamma^{-3-1}$, and $k=\Gamma^{-1-2}$. Three sets of complex conjugation operators are given by $\Sigma_{z,1} = \Gamma^{4-2}$, $\Sigma_{z,2} = \Gamma^{4-3}$, and $\Sigma_{z,3} = \Gamma^{4-1}$ \cite{Chester:2020nim}. Therefore, complex structure can be found from the real Majorana-Weyl spinor. This should not be surprising, as ${\bf128}$ corresponds to $(\mathbb{O}_s\otimes\mathbb{O})\mathbb{P}^2$, where $\mathbb{H} \subset \mathbb{O}_s$. Recall that $\mathbb{O}_s$ gives $Spin(4,4)$ rotations, which was motivated by Kostant to have utility for three generations. The following action provides a concrete realization of this idea
\begin{eqnarray}
S &=& \int d^4x \left(\bar{\psi}_1 i \Gamma^\mu \partial_\mu \psi_1 + \bar{\psi}_2 j \Gamma^\mu \partial_\mu \psi_2 + \bar{\psi}_3 k \Gamma^\mu \partial_\mu \psi_3 \right) \nonumber  \\
&=& \int d^4x \left(\left(\Sigma_{z,1} \psi_1\right)^\top \Gamma^0 \Gamma^{-2-3}\Gamma^\mu \partial_\mu \psi_1 + \left(\Sigma_{z,2} \psi_2\right)^\top \Gamma^0 \Gamma^{-3-1}\Gamma^\mu \partial_\mu \psi_2 \right. \nonumber \\
&& \left. + \left(\Sigma_{z,3} \psi_3\right)^\top \Gamma^0 \Gamma^{-1-2}\Gamma^\mu \partial_\mu \psi_3\right) \nonumber \\
&=& \int d^4x \Psi^\top \left(\sum_{i=1}^3 P_i^N \Sigma_{z,i}\Gamma^0 \Gamma^{-i} \Gamma^{-3-2-1}\Gamma^\mu \partial_\mu \right) \Psi \label{3gens}
\end{eqnarray}
where $\Sigma_{z,i}^\top = \Sigma_{z,i}$ and $(P_i^N)^\top = P_i^N$.

A gauge theory with $\mathfrak{su}_{2,2}\oplus \mathfrak{u}_1 \oplus \mathfrak{su}_5$ symmetry can be introduced, with the matter fields being in 1:1 correspondence with the generators living in $\Gamma(\mathfrak{e}_{8(-24)}) \ominus  (\mathfrak{su}_{2,2}\oplus \mathfrak{u}_1 \oplus \mathfrak{su}_5)$ as the matter sector. Provided that the appropriate Higgs sector can be identified, spin-1 gauge bosons for conformal gravity and flipped $SU_{5}$ GUT are found with three generations of fermions of the standard model.

\section{Fields and action of the flipped conformal GUT theory}
\label{theory}

The new theory presented contains 40 vector gauge bosons from $SU_{2,2}\times U_{1} \times SU_{5}$ and 208 matter fields, giving 368 real off-shell degrees of freedom. Following Ivanov and Niederle's construction of gauge gravity \cite{Ivanov1982}, the matter sector is considered as a scalar with respect to the global manifold, but the fermions obtain their spin as representations of the gauge group $SL_{2}(\mathbb{C})_{\mathbb{R}}\subset SU_{2,2}$.

The branching rules of $\Gamma(\mathfrak{e}_{8(-24)}) \rightarrow \Gamma(\mathfrak{su}_{3,2})\oplus \mathfrak{su}_5 \rightarrow \mathfrak{su}_{2,2} \oplus \mathfrak{u}_1 \oplus \mathfrak{su}_5 \rightarrow \mathfrak{sl}_{2,\mathbb{C}} \oplus \mathbb{R} \oplus \mathfrak{u}_1 \oplus \mathfrak{su}_5$ identify the field content for a chart,
\begin{eqnarray}
\Gamma(\mathfrak{su}_{3,2})\oplus \mathfrak{su}_5 &\rightarrow& \mathfrak{su}_{2,2}\oplus \mathfrak{u}_1 \oplus \mathfrak{su}_5 \rightarrow \mathfrak{sl}_{2,\mathbb{C}} \oplus \mathbb{R} \oplus \mathfrak{u}_1 \oplus \mathfrak{su}_5, \\
({\bf 24},{\bf1}) &=& \underbrace{({\bf15},{\bf1})_0}_{\omega_{\mu\,\,\,\bar{b}}^{\,\,\,\bar{a}}} \oplus \underbrace{({\bf1},{\bf1})_0}_{a_\mu} \oplus \underbrace{({\bf4},{\bf1})_{5} \oplus (\overline{{\bf4}},{\bf1})_{-5}}_{\psi^{\bar{\alpha}} + \bar{\psi}_{\bar{\alpha}}}, \nonumber \\
&=& \underbrace{({\bf3},{\bf1},{\bf1})_{0,0} \oplus ({\bf1},{\bf3},{\bf1})_{0,0}}_{\omega_{\mu\,\,\, b}^{\,\,\, a}} \oplus \underbrace{({\bf1},{\bf1},{\bf1})_{0,0}}_{d_\mu} \oplus  \underbrace{({\bf2},{\bf2},{\bf1})_2}_{e_\mu^a} \oplus \underbrace{({\bf2},{\bf2},{\bf1})_{-2}}_{ f_\mu^a} \nonumber \\
&& \oplus \underbrace{({\bf1},{\bf1},{\bf1})_{0,0}}_{a_\mu} \oplus \underbrace{({\bf2},{\bf1},{\bf1})_{1,5} \oplus ({\bf1},{\bf2},{\bf1})_{1,-5}}_{\psi^{\alpha} + \bar{\psi}_{\dot{\alpha}}} \oplus \underbrace{({\bf1},{\bf2},{\bf1})_{-1,5} \oplus ({\bf2},{\bf1},{\bf1})_{-1,-5}}_{\psi_M^{\alpha} + \bar{\psi}_{M\dot{\alpha}}}, \nonumber \\
({\bf5},{\bf10}) &=& \underbrace{({\bf4},{\bf10})_1}_{\lambda^{\bar{\alpha}\mathring{\imath}\mathring{\jmath}}} \oplus \underbrace{({\bf1},{\bf10})_{-4}}_{H^{\mathring{\imath}\mathring{\jmath}}} \nonumber \\
&=& \underbrace{({\bf2},{\bf1},{\bf10})_{1,1}}_{\lambda^{\alpha\mathring{\imath}\mathring{\jmath}}} \oplus \underbrace{({\bf1},{\bf2},{\bf10})_{-1,1}}_{\lambda_{M\dot{\alpha}}^{\,\,\,\,\,\,\mathring{\imath}\mathring{\jmath}}} \oplus \underbrace{({\bf1},{\bf1},{\bf10})_{0,1}}_{H^{\mathring{\imath}\mathring{\jmath}}}, \nonumber
\end{eqnarray}
and
\begin{eqnarray}
({\bf10},\overline{{\bf5}}) &=& \underbrace{({\bf6},\overline{{\bf5}})_{2}}_{h^{\bar{a}}_{\,\,\,\mathring{\imath}}} \oplus \underbrace{({\bf4}, \overline{{\bf5}})_{-3}}_{\chi^{\bar{\alpha}}_{\,\,\,\mathring{\imath}}} \\
&=& \underbrace{({\bf2},{\bf2},\overline{{\bf5}})_{0,2}}_{g^a_{\,\,\,\mathring{\imath}}} \oplus \underbrace{({\bf1},{\bf1},\overline{{\bf5}})_{2,2}}_{h_{\mathring{\imath}}} \oplus \underbrace{({\bf1},{\bf1},\overline{{\bf5}})_{-2,2}}_{h_{M\mathring{\imath}}} \oplus \underbrace{({\bf2},{\bf1},\overline{{\bf5}})_{1,-3}}_{\chi^{\alpha}_{\,\,\,\mathring{\imath}}} \oplus \underbrace{({\bf1},{\bf2},\overline{{\bf5}})_{-1,-3}}_{\chi_{M\dot{\alpha}\mathring{\imath}}}. \nonumber
\end{eqnarray}
The indices run over $\mathring{\imath} = 1, 2,\dots 5$ for $\mathfrak{su}_5$, $\bar{\alpha}=1,2,3, 4$ and $\bar{a} = 1,2,\dots 6$ for $SU(2,2)$, and $\alpha, \dot{\alpha} = 1,2$ and $a=1, 2, 3, 4$ for $\mathfrak{sl}_{2,\mathbb{C}}$. The projection operators in the fermionic action remove the spinors $\psi_M^\alpha$ and $\chi_M{\dot{\alpha}}\mathring{\imath}$. This allows for three generations of matter fields $\lambda^{i\alpha\mathring{\imath}\mathring{\jmath}}$, $\chi^{i\alpha}_{\ \mathring{i}}$, and $\psi^{i\alpha}$, which stem from the three ${\bf16}$ spinors of $Spin_{10}$ found in Eq.~\eqref{3gens}.

The action $S = S_{gauge} + S_{kin} + S_{pot}$ contains all of the gauge fields in $S_{gauge}$, the kinetic matter terms in $S_{kin}$, and potential matter terms in $S_{pot}$. The gauge bosons $\omega_{\mu \,\,\,\ \bar{b}}^{\,\,\, \bar{a}}$, $a_mu$, and $A_{\mu\,\,\, \mathring{\jmath}}^{\,\,\,\mathring{\imath}}$ lead to field strengths $R_{\mu\nu\,\,\, \bar{b}}^{\,\,\,\,\,\,\bar{a}}$, $f_{\mu\nu}$, and $F_{\mu\nu\,\,\,\mathring{\jmath}}^{\,\,\,\,\,\,\mathring{\imath}}$ for $SU(2,2)$, $U(1)$, and $SU(5)$, respectively. Using differential forms such as $f = \frac{1}{2}f_{\mu\nu} dx^\mu \wedge dx^\nu$, the action for the gauge bosons is
\begin{equation}
    S_{gauge} = \int \left(\alpha |e| \left(\mathcal{R}\wedge \mathcal{R} - R \wedge R\right) +\beta \mbox{tr}\left(\mathcal{R}\wedge *\mathcal{R} - R \wedge *R\right) + \frac{1}{g^2_5} \mbox{tr}\left(F \wedge * F\right) + \frac{1}{g^2_X} f \wedge * f \right),  \label{Sgauge}
\end{equation}
where $\alpha = 3c^3/16\pi G \Lambda$ and $\beta = -3c^3/64\pi G \Lambda$.

Historically, Kaku, Townsend, and van Nieuwenhuizen found conformal gauge gravity by generalizing the MacDowell-Mansouri action to the superconformal group, but were not able to recover the Einstein-Hilbert action \cite{Kaku:1977pa}. Ivanov and Niederle successfully showed how to obtain Einstein gravity \cite{Ivanov:1981wm,Ivanov:1984nu}. Gauging $Spin_{3,2}$ or $Spin_{4,1}$ leads to the same theory for the MacDowell-Mansouri and Yang-Mills actions when the $Spin_{3,1}$ piece is subtracted away. However, these differ for $SU_{2,2}$, which allows for a linear combination of MacDowell-Mansouri gravity and Yang-Mills gravity to be found in a unique way to lead to the Einstein-Hilbert action with cosmological constant. It will be proven elsewhere that the combination of $\alpha$ and $\beta$ in Eq.~\eqref{Sgauge} leads to the Einstein-Hilbert action with cosmological constant when applying a conformal transformation to uncover a non-propagating scalar field.

The ``kinetic'' terms for the matter fields are generalized to contain some interactions via the covariant derivatives
\begin{eqnarray}
S_{kin} &=& \int d^4x |e| \mathcal{L}_{kin}= S_{kin,\lambda} + S_{kin,\chi} + S_{kin,\psi} + S_{kin,H} + S_{kin,h} + S_{kin,g}, \\
\mathcal{L}_{kin,\lambda} &=& -i \lambda^\dagger_{i\mathring{\imath}\mathring{\jmath}} \bar{\sigma}^a e^\mu_a \overleftrightarrow{D}_\mu \lambda^{i\mathring{\imath}\mathring{\jmath}} = -\frac{i}{2}\left(\lambda^\dagger_{i\mathring{\imath}\mathring{\jmath}} \bar{\sigma}^a e^\mu_a (D_\mu \lambda)^{i\mathring{\imath}\mathring{\jmath}}  -(D_\mu \lambda^\dagger)_{i\mathring{\imath}\mathring{\jmath}} \bar{\sigma}^a e^\mu_a \lambda^{i\mathring{\imath}\mathring{\jmath}}\right), \\
\mathcal{L}_{kin,\chi} &=& -i \chi^{\dagger\,\,\,\mathring{\imath}}_{\,\,\, i} \bar{\sigma}^a e^\mu_a \overleftrightarrow{D}_\mu \chi^i_{\,\,\,\mathring{\imath}} = -\frac{i}{2}\left(\chi^{\dagger \,\,\,\mathring{\imath}}_{\,\,\, i} \bar{\sigma}^a e^\mu_a (D_\mu \chi)_{\,\,\,\mathring{\imath}}^i  -(D_\mu \chi^\dagger)^{\,\,\,\mathring{\imath}}_i \bar{\sigma}^a e^\mu_a \chi_{\,\,\,\mathring{\imath}}^i\right), \\
\mathcal{L}_{kin,\psi} &=& -i \psi^{\dagger}_i \bar{\sigma}^a e^\mu_a \overleftrightarrow{D}_\mu \psi^i = -\frac{i}{2}\left(\psi^{\dagger}_{i} \bar{\sigma}^a e^\mu_a D_\mu \psi^i  -D_\mu \psi^{\dagger}_i \bar{\sigma}^a e^\mu_a \psi^i\right), \\
\mathcal{L}_{kin,H} &=& -\left(D_\mu H\right)^{\mathring{\imath}\mathring{\jmath}} \left(D^\mu H \right)^\dagger_{\mathring{\imath}\mathring{\jmath}} , \\
\mathcal{L}_{kin,h} &=& -\left(D_\mu h\right)_{\mathring{\imath}} \left(D^\mu h \right)^{\dagger\mathring{\imath}}, \\
\mathcal{L}_{kin,g} &=& - \frac{1}{2}e_\mu^a(D_a g_b^{\mathring{\imath}} - D_b g_a^{\mathring{\imath}})e^\mu_c(D^c g^{\dagger b}_{\mathring{\imath}} - D^b g^{\dagger c}_{\mathring{i}}).
\end{eqnarray}
The one field that is unique to this theory is the complex vector Higgs of $SU_{5}$ with $U_{1}$ charge $g^a_{\mathring{\imath}}$.

The covariant derivatives are provided by the following,
\begin{eqnarray}
 (D_\mu \lambda)^{i\alpha \mathring{\imath}\mathring{\jmath}} &=& \partial_\mu \lambda^{i\alpha \mathring{\imath}\mathring{\jmath}} + \omega_{\mu \,\,\,\beta}^{\,\,\,\alpha} \lambda^{i\beta \mathring{\imath}\mathring{\jmath}} -i A_{\mu\,\,\, \mathring{k}}^{\,\,\,\mathring{\imath}} \lambda^{i\alpha \mathring{k}\mathring{\jmath}} -i A_{\mu\,\,\, \mathring{k}}^{\,\,\,\mathring{\jmath}} \lambda^{i\alpha \mathring{\imath}\mathring{k}} - ia_\mu \lambda^{i\alpha \mathring{\imath}\mathring{\jmath}}, \\
 (D_\mu \chi)_{\,\,\,\,\,\,\mathring{\imath}}^{i\alpha} &=& \partial_\mu \chi_{\,\,\,\,\,\,\mathring{\imath}}^{i\alpha} + \omega_{\mu \,\,\,\beta}^{\,\,\,\alpha}\chi_{\,\,\,\,\,\,\mathring{\imath}}^{i\beta} + i A_{\mu\,\,\,\mathring{\imath}}^{\,\,\,\mathring{\jmath}} \chi_{\,\,\,\,\,\,\mathring{\jmath}}^{i\alpha} +3 ia_\mu \chi_{\,\,\,\,\,\,\mathring{\imath}}^{i\alpha}, \\
 (D_\mu \psi)^{i\alpha} &=& \partial_\mu \psi^{i\alpha} + \omega_{\mu \,\,\,\beta}^{\,\,\,\alpha}\psi^{i\beta} - 5 i a_\mu \psi^{i\alpha}, \\
 (D_\mu H)^{\mathring{\imath}\mathring{\jmath}} &=& \partial_\mu H^{\mathring{\imath}\mathring{\jmath}} - i A_{\mu\,\,\,\mathring{k}}^{\,\,\,\mathring{\imath}} H^{\mathring{k}\mathring{\jmath}}- i A_{\mu\,\,\,\mathring{k}}^{\,\,\,\mathring{\jmath}} H^{\mathring{\imath}\mathring{k}} +4i a_\mu H^{\mathring{\imath}\mathring{\jmath}}, \\
 (D_\mu h)_{\mathring{\imath}} &=& \partial_\mu h_{\mathring{\imath}} +i A_{\mu\,\,\,\mathring{\imath}}^{\,\,\,\mathring{\jmath}}h_{\mathring{\jmath}} -2 i a_\mu h_{\mathring{\imath}}, \\
 (D_\mu g)^{a}_{\,\,\,\mathring{\imath}} &=& \partial_\mu g^{a}_{\,\,\,\mathring{\imath}} + \omega_{\mu \,\,\,b}^{\,\,\,a}g^{b}_{\,\,\,\mathring{\imath}} +i A_{\mu\,\,\,\mathring{\imath}}^{\,\,\,\mathring{\jmath}} g^a_{\,\,\,\mathring{\jmath}} - 2 i a_\mu g^{a}_{\,\,\,\mathring{\imath}}.
\end{eqnarray}
where the gravitational sector only requires the Lorentzian spin connection, since conformal gauge gravity allows for the gauge fields of the coset $SU_{2,2}/SL_{2}(\mathbb{C})_{\mathbb{R}}\simeq Spin_{4,2}/Spin_{3,1}$ to be solved for \cite{Kaku:1977pa}. %The gauge field of the scale transformation will be shown elsewhere to

The Higgs and Yukawa potentials found in the matter sector are given by all possible terms up to quartic order, which contains the Higgs fields found in flipped $SU_{5}$ SUSY GUT, but does not use a cubic superpotential \cite{ANTONIADIS1987, Ellis:1993ks},
\begin{eqnarray}
S_{pot} &=& \int d^4x |e| \mathcal{L}_{pot} = S_{pot,H} + S_{pot,h} + S_{pot,g} + S_{pot,Hh} + S_{pot,Hg} + S_{pot,hg} + S_{Yukawa},  \nonumber\\
\mathcal{L}_{pot,H} &=& \frac{\mu_H^2}{2} |H|^2 - \frac{\lambda_{H,1}}{4}|H|^4 - \frac{\lambda_{H,2}}{4} |H|^{2\mathring{\imath}}_{\,\,\,\,\,\,\mathring{\jmath}} |H|^{2\mathring{\jmath}}_{\,\,\,\,\,\,\mathring{\imath}} , \\
\mathcal{L}_{pot,h} &=& \frac{\mu_h^2}{2}|h|^2 -\frac{\lambda_{h}}{4}|h|^4 , \\
\mathcal{L}_{pot,g} &=& \frac{\mu_g^2}{2}|g|^2 - \frac{\lambda_{g,1}}{4}|g|^4 - \frac{\lambda_{g,2}}{4}|g|^{2a}_{\,\,\,\,\,\, b} |g|^{2b}_{\,\,\,\,\,\, a} , \label{potg} \\
\mathcal{L}_{pot,Hh} &=& - \frac{\lambda_{Hh,1}}{4}|H|^2|h|^2 -\left( \frac{\lambda_{Hh,2}}{3!}\epsilon_{\mathring{\imath}\mathring{\jmath}\mathring{k}\mathring{l}\mathring{m}}H^{\mathring{\imath}\mathring{\jmath}} H^{\mathring{k}\mathring{l}} h^{\dagger \mathring{m}} + h.c. \right), \\
\mathcal{L}_{pot,Hg} &=& - \frac{\lambda_{Hg,1}}{4}|H|^2|g|^2  - \frac{\lambda_{Hg,2}}{4} g^a_{\mathring{\imath}}g^{\dagger\mathring{\jmath}}_a H^\dagger_{\mathring{\jmath}\mathring{k}}H^{\mathring{k}\mathring{\imath}}, \\ %- \frac{\lambda_{Hg,2}}{3!}Hg^a_{\mathring{\imath}}g^{\dagger\mathring{\imath}}_a H, \\
\mathcal{L}_{pot,hg} &=& - \frac{\lambda_{hg,1}}{4}|h|^2|g|^2 - \frac{\lambda_{hg,2}}{4}h_{\mathring{\imath}} h^{\dagger \mathring{\jmath}} g^a_{\mathring{\jmath}} g^{\dagger\mathring{\imath}}_a, \\
\mathcal{L}_{Yukawa} &=& Y_d^{ij}\lambda_i^{\alpha\mathring{\imath}\mathring{\jmath}} h^{\dagger \mathring{k}} \lambda_{j \alpha}^{\mathring{l}\mathring{m}}\epsilon_{\mathring{\imath}\mathring{\jmath}\mathring{k}\mathring{l}\mathring{m}} + Y_{u\nu}^{ij}\lambda_i^{\alpha\mathring{\imath}\mathring{\jmath}}h_{\mathring{\imath}}\chi_{j\alpha\mathring{\jmath}} + Y_e^{ij} \chi_{i}^{\alpha\mathring{\imath}}h_{\mathring{\imath}}\psi_{\alpha} + h.c. , \label{yukawa} % \\
%&& + Y_{\nu}^{ij} \lambda_i^{\alpha\mathring{\imath}\mathring{\jmath}} g_{\mathring{\imath}}^a \sigma_{a\alpha}^{\,\,\,\,\,\,\beta}\chi^\dagger_{j\dot{\beta}\mathring{\jmath}} + Y_{u}^{ij} \lambda_{i\dot{\alpha}\mathring{\imath}\mathring{\jmath}}^{\dagger} g^{\dagger \mathring{\imath}}_a \bar{\sigma}^{a\dot{\alpha}}_{\,\,\,\,\,\,\beta}\chi_{j\beta\mathring{\jmath}} + h.c. , \nonumber
\end{eqnarray}
%  +Y_\nu^{ij} \lambda_i^{\alpha\mathring{\imath}\mathring{\jmath}}H^\dagger_{\mathring{\imath}\mathring{\jmath}}H^{\mathring{k}\mathring{l}}h_{\mathring{k}} \chi_{j\alpha \mathring{l}}
where $|H|^2 = H^{\mathring{\imath}\mathring{\jmath}}H^\dagger_{\mathring{\imath}\mathring{\jmath}}$ and $|H|^{2\mathring{\imath}}_{\,\,\,\,\,\,\mathring{\jmath}} = H^{\mathring{\imath}\mathring{k}}H^\dagger_{\mathring{k}\mathring{\jmath}}$. It can be shown that this contains mass terms for the three families of down quarks, up quarks, neutrinos, and electrons.

\section{Conclusions}
\label{conc}

In conclusion, we have realized the $\mathcal{N}=1$ superquasiconformal Lie algebra in $D=10+1$ space-time dimensions as the Grassmann envelope of the simple, minimally non-compact real Lie algebra $\mathfrak{e}_{8(-24)}$ : $\Gamma \left( \mathfrak{e}_{8(-24)}\right) \simeq \mathfrak{sqconf}\left(
\mathbb{R}\oplus J_{2}(\mathbb{O})\right) $. Within this framework, the appropriate use of projection operators allowed us to introduce an action with three generations of standard model fermions from a unique Majorana-Weyl ${\bf128}$ semispinor irrepr. of $Spin_{12,4}$, by identifying a quaternionic structure with extra time dimensions for mass/flavor oscillations. Then, a gauge theory with gauge group $G=SU_{2,2}\times U_{1}\times SU_{5}$ was presented, with the matter sector in 1:1 correspondence with the fields coordinatizing the (super)coset $\Gamma \left( E_{8(-24)}\right) /G$, where $\Gamma \left( E_{8(-24)}\right) $ is the Lie supergroup whose Lie superalgebra is $\Gamma \left( \mathfrak{e}_{8(-24)}\right) $. As an interesting outcome, we found a quite natural realization of the (non-maximal) embedding of the
$\mathcal{N}=1$ superconformal algebra in $3+1$ dimensions into the $\mathcal{N}=1$
superquasiconformal Lie algebra in $10+1$ dimensions : $\mathfrak{sconf}\left( J_{2}(\mathbb{C})\right) \subset \mathfrak{sqconf}%
\left( \mathbb{R}\oplus J_{2}(\mathbb{O})\right) $. In forthcoming works, we plan to deal with a mathematical treatment of the Grassmann envelope and its real forms, as well as with an extended and detailed treatment of the gravity theory presented here.

\end{document}